# Cathodoluminescence Wavefront Retrieval

*Izzah Machfuudzoh[1,*], Ryoichi Horisaki[2], Takumi Sannomiya[1,**]*

[1] Department of Materials Science and Engineering, School of Materials and Chemical Technology, Institute of Science Tokyo, 4259 Nagatsuta, Midori-ku, Yokohama 226-8503 Japan

[2] Department of Information Physics and Computing, Graduate School of Information Science and Technology, The University of Tokyo, 7-3-1 Hongo, Bunkyo-ku, Tokyo 113-8656 Japan

**Corresponding authors**

* Izzah Machfuudzoh (Email: machfuudzoh.i.aa@m.titech.ac.jp)

** Takumi Sannomiya (Email: sannomiya@mct.isct.ac.jp)

**Abstract**

Free-electron-based nanoscopy enables the study of optical excitations in materials with deep-subwavelength spatial resolution, with cathodoluminescence (CL) being one of the most useful resulting signals. When combined with an optical collection system, CL measurements can access multidimensional information of light; yet the phase of the emitted optical fields has remained largely elusive. Here, we demonstrate a reference-free phase retrieval approach for far-field CL wavefronts using the Gerchberg-Saxton algorithm implemented with real-space and angular-space CL intensity data. Applying this approach to representative nanostructures, including a planar surface, nanosphere, plasmonic crystal, and nanowire, we reconstruct distinct phase distributions that reveal their underlying radiation mechanisms. This reference-free framework offers a robust and flexible route for retrieving the phase of electron-beam-excited optical fields without relying on a reference wave, making it readily extendable to a wide range of nanostructures.

Cathodoluminescence (CL) spectroscopy is a powerful technique for probing optical excitations in nanostructures with nanometer spatial resolution using a focused electron beam. The time-varying evanescent electric field around the electron trajectory polarizes the material, producing coherent emissions such as transition radiation and surface-plasmon-mediated light[1-3]. When combined with a parabolic mirror collection system, CL measurements can provide a wide range of information about the emitted radiation as a function of excitation position, including spatial analysis, spectral response, angular distribution, and polarimetry[4-7]. These capabilities have enabled investigations of radiation mechanisms in plasmonic nanoantennas[4, 8], band structures and Bloch modes in photonic crystals[9, 10], and field distributions of multipole modes in nanospheres[11, 12]. Recently developed CL techniques for the direct visualization of photon emission in real space further enable the study of the electromagnetic energy transport in plasmonic materials[13-15], including approaches based on indirect methods that leverage the interference of coherent channels to estimate the emission positions[16, 17].

Despite the rich multidimensional information provided by CL techniques, conventional measurements primarily offer access only to intensity distributions, while the phase of the emitted optical field remains largely elusive. This phase itself plays a critical role in determining the structure of scattered optical wavefronts, which ultimately defines the full complex electromagnetic field emitted by nanoscale optical sources. Several attempts have been made to access phase information through interference mechanisms by utilizing transition radiation as a coaxial reference background[18-20]. Another study reported leveraging holographic configurations to probe surface plasmon polariton (SPP) scattering phases; however, these are bound to secondary propagation and scattering phenomena tailored by specific, engineered sample boundaries[18], rather than capturing the direct, intrinsic CL generation phase at the localized point of electron impact. Furthermore, these methods require careful balancing of the relative intensities of coherent channels, precise geometric positioning relative to the scatterers[18, 21], and strict mutual coherence among the different emission sources involved[22] to resolve stable interference fringes. This heavy reliance on geometrically specialized sample configurations, as well as the fact that the retrieved phase is that of the scattered field excited by SPPs, limits their general practicality across complex, unpatterned nanostructures. These limitations highlight the critical need for a "true" CL-based phase-retrieval method that operates independently of any physical reference fields or spatial boundary conditions, while still enabling accurate reconstruction of the true CL emission wavefront using standard, experimentally accessible intensity measurements.

In this work, we demonstrate reference-free CL wavefront retrieval via real- and reciprocal-space CL imaging based on the error-reduction Gerchberg-Saxton algorithm[23]. By applying an iterative computational algorithm to the dataset of experimentally measured emission images in real space and angular images in reciprocal angular space, the phase of the optical CL wavefront can be reconstructed without requiring interferometric detection. The method is applied to a range of nanostructures, including a planar surface, nanosphere, plasmonic crystal, and nanowire, each exhibiting distinct radiation mechanisms and corresponding relative phase distributions. This CL-based reference-free phase-retrieval approach provides direct access to the far-field optical CL wavefronts of nanophotonic structures, offering valuable insights for engineering optical wavefronts in light-based applications such as integrated optics[24] and optical computing[25]. We note that the phase retrieved in this work is attributed to the coherent CL component, which produces a deterministic complex field, although this approach is not limited by the origin of the radiation.

The phase retrieval of optical CL wavefronts was performed using an iterative Gerchberg-Saxton algorithm[23], alternating between two Fourier-transform-related domains, namely real space and angular space, with datasets for each domains acquired using a modified scanning transmission electron microscope (STEM)[13, 26] (Figure 1(a)). The input datasets consist of three pixelated images (Figure 1(d)): a randomly generated initial phase within the range [-π, π] (image 1), and the amplitudes (square roots of the intensity) of the real-space emission position image (image 2; emission spot(s) in the $x'y'$-plane of real space (Figure 1(c))) and of the angular image (image 3; angular emission pattern representing angular space projected onto the $YZ$-plane (Figure 1(c))). Both intensity images were acquired using an emission-position imaging system with an angle-resolving pinhole mask positioned at $\theta = 20°$ and $\phi = 0°$ (Figure 1(c)). The polar angle $\theta = 20°$ was selected because it captures a large portion of the upward emission pattern within its 2.4 sr detection solid angle.

The algorithm starts in real space by combining the initial random phase (image 1) with the emission image amplitude (image 2). The resulting wave function is Fourier transformed into angular space, yielding the retrieved angular amplitude (image 4) and phase distribution (image 5). An amplitude constraint is then applied by replacing the retrieved amplitude (image 4) with the acquired amplitude (image 3), while preserving the retrieved phase (image 5), thereby updating the angular-space wave function. The updated angular-space wave function is then inverse Fourier transformed back to real space. The retrieved emission amplitude (image 6) is subjected to an amplitude constraint and replaced with the acquired emission amplitude (image 2), while preserving the retrieved real-space phase (image 7). The updated real-space wave function is subsequently Fourier transformed back into angular space.

This iterative process is repeated until the mean squared error (MSE) between the acquired amplitudes (images 2 and 3) and the reconstructed amplitudes (images 4 and 6) approaches zero in their respective domains. The algorithm ultimately reconstructs the phase distributions in both real and angular spaces. In this study, we focus on the angular-space phase distribution (image 5*) because it directly represents the spatial phase of the far-field optical CL wavefronts at each angular coordinate ($\theta$, $\phi$), with the phase differences between different angular components representing the relative phase of the wavefront. Therefore, what we obtain is a "relative phase" distribution within the obtained radiation image rather than a global phase with respect to some reference.

Prior to being used as input for the algorithm, the emission and angular images were preprocessed and resampled to ensure accurate correspondence between the two Fourier-transform-related domains. See Figure 1(d) for the workflow of the phase-retrieval algorithm and the corresponding image numbers. Refer also to the Supporting Information (SI) for an evaluation of the algorithm's performance. Additionally, three-dimensional (3D) datasets of spatial CL photon maps, each containing a CL spectrum at every pixel, were acquired using an angle-resolved spectral system for optical mode analysis (Figure 1(b)). See SI for details of the CL measurements.

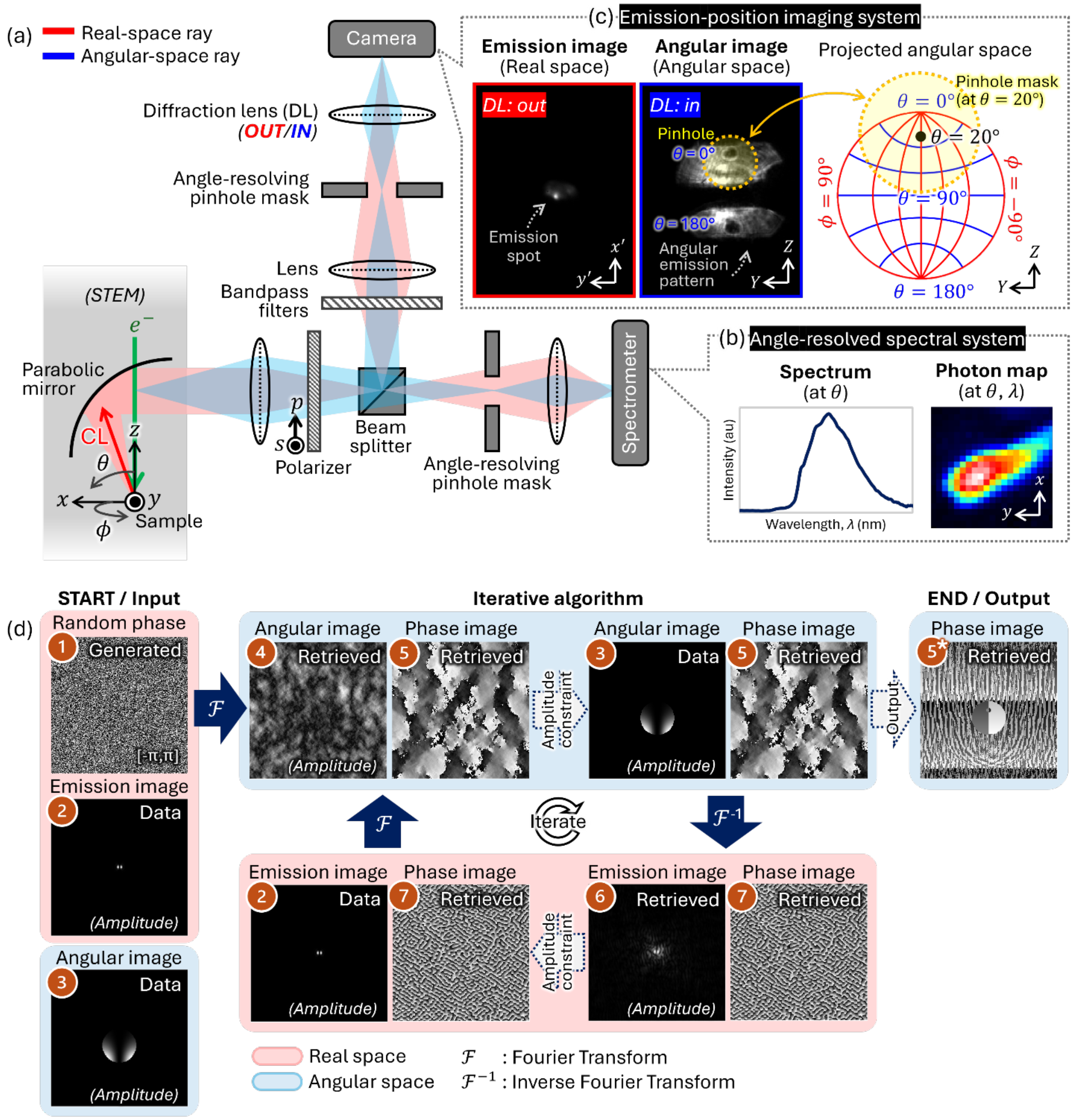

**Figure 1**. Experimental setup and phase-retrieval algorithm. (**a**) Optical ray paths of the CL-based data acquisition system. (**b**) One optical path leads to an angle-resolved spectral system that acquires angle- (wavenumber or momentum) and wavelength-resolved CL photon maps and their spectra. (**c**) The second optical path leads to an emission-position imaging system that acquires an emission image in real space and an angular image in angular space by focusing either the real-space rays

(red) or the angular-space rays (blue) onto the camera, sequentially, by removing or inserting a diffraction lens (DL), respectively. The projection of angular space ($\theta$, $\phi$) onto the *YZ*-plane is also shown, together with the angle-resolving pinhole mask indicated by yellow circles. Both systems employed identical pinhole mask configurations with a detection solid angle of 2.4 sr positioned at an emission angle of ($\theta$, $\phi$) = (20°, 0°). (**d**) Flow diagram of the phase retrieval based on the Gerchberg-Saxton algorithm. The red and blue regions denote real and angular spaces, respectively. The images are labelled (random phase, emission image, angular image, and phase image), numbered (1 to 7), and marked according to their origin (numerically generated, experimentally acquired, and algorithmically retrieved) to facilitate discussion. Additional annotations (phase range [-π, π], amplitude correspondence, Fourier transform $F$, and inverse Fourier transform $F^{-1}$) are also indicated. The algorithm was developed and implemented in Matlab.

The phase of the optical CL wavefronts originating from transition radiation (TR) emitted by a planar silver surface was retrieved. The surface consists of an approximately 200 nm thick vapor-deposited silver film capped with a 5 nm $SiO_2$ layer to prevent surface degradation (backscattered electron image in Figure 2(c)). The emitted CL was collected through a pinhole mask with a 2.4 sr solid angle at ($\theta$, $\phi$) = (20°, 0°) and filtered using a *p*-polarizer to select optical electric field components oscillating in the *xz*-plane (Figure 2(a)). The polarized CL spectrum at $\theta$ = 20°, integrated over the entire sample area (inset), is presented in Figure 2(b). By raster scanning the electron beam across the surface, a corresponding CL photon map integrated over 400-700 nm (gray region in Figure 2(b)) is shown in Figure 2(d). The uniform bright contrast across the surface indicates homogeneous excitation of TR by high-energy electrons[1]. For comparison, simulations based on the point spread function under the paraxial approximation[13, 14] were performed using the same experimental settings with a photon wavelength of $\lambda$ = 500 nm.

For phase-retrieval data acquisition, the electron beam was positioned at the center of the sample scanning area (red dot in Figure 2(c)) to minimize phase ramps in angular space (see the SI for details). The corresponding experimental emission and angular intensity images are shown in Figure 2(e, f), respectively. The single emission spot in Figure 2(e) agrees well with the simulated result in Figure 2(h), obtained by modeling TR as radiation from an out-of-plane electric dipole oriented along the $z$-axis. A similar agreement is observed in the angular images (Figure 2(f, i)), where the emission pattern is concentrated within the polar angle range of approximately $30° \leq \theta \leq 80°$, consistent with the expected dipolar angular radiation pattern of TR[21, 27]. The upper half of the projected angular coordinates is overlaid onto the angular-space images for reference.

The retrieved phase distribution corresponding to the intensity data in Figure 2(e, f) is shown in Figure 2(g). The phase exhibits a nearly homogeneous distribution within the angular region $30° \leq \theta \leq 80°$, where the TR angular intensity is present and the phase information is physically meaningful. This result is in good agreement with the simulated phase distribution shown in Figure 2(j). For quantitative analysis, line profiles extracted along the directions indicated by the arrows in Figure 2(g, j) show that the retrieved relative phase $\Delta\psi$ is approximately 0 rad (green curve in Figure 2(k)), consistent with the simulated relative phase (orange curve). These results indicate that TR produces an in-phase wavefront with a uniform phase distribution across angular space, supporting its established role as a reference field in the far-field region for various interferometric studies[16, 18, 20, 22].

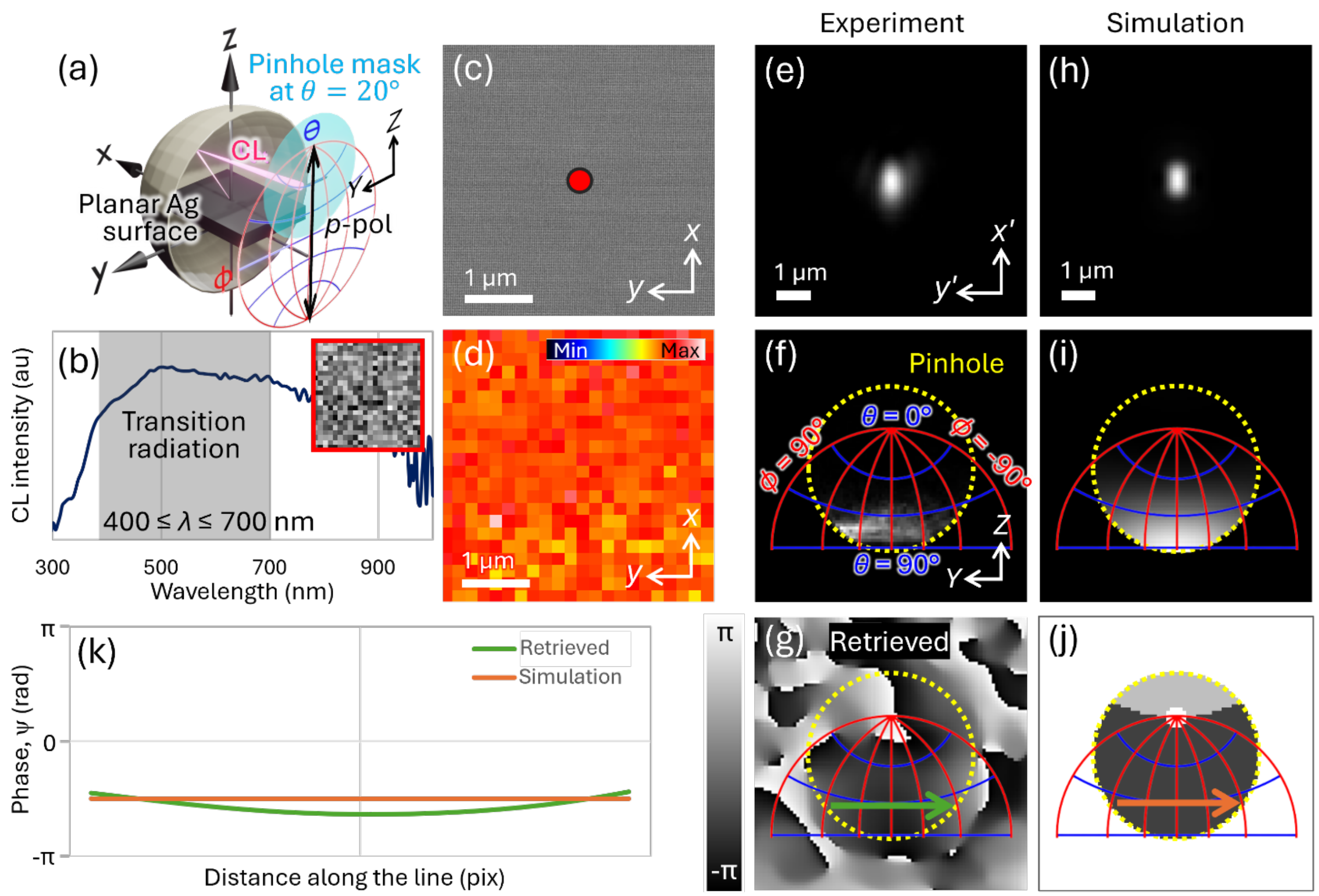

**Figure 2**. Transition radiation from a planar silver surface. (**a**) Schematic illustration of the planar surface inside a parabolic mirror. The emitted CL is selected by a pinhole mask with a detection solid angle of 2.4 sr located at ($\theta$, $\phi$) = (20°, 0°), indicated by the blue circle, and filtered by a *p*-polarizer (black arrow) aligned along the *xz*-plane. (**b**) CL spectrum measured at $\theta = 20°$, integrated over the entire surface area shown in the inset. The gray rectangle indicates the selected wavelength range of 400 nm ≤ λ ≤ 700 nm. (**c**) Backscattered electron image of the planar silver surface. The red dot indicates the electron beam position. (**d**) CL photon map at $\theta$ = 20°, integrated over the wavelength range indicated in (b). Experimental (**e**) emission image and (**f**) angular image acquired at the beam position shown in (c) within the wavelength range indicated in (b), together with the corresponding (**g**) retrieved phase image. Simulated (**h**) emission image, (**i**) angular image, and (**j**) phase image. (**k**) Line profiles of the phase distribution extracted along the directions of the arrows

in (g, j). The angular-space images in (f, g, i, j) also show the pinhole mask (yellow circle) and the upper half of the projected angular coordinates. The images in (e-j) are magnified by about ×3.1 relative to the original size used in the retrieval process.

The phase of the optical CL wavefronts emitted by an in-plane electric dipole from a 140 nm silver nanosphere (STEM bright-field image in Figure 3(c)) was retrieved, using $s$-polarization along the $y$-axis (Figure 3(a)). The silver spheres were fabricated by vapor deposition in an argon atmosphere at 1400 Pa[6, 12]. Silver was selected because its sharp electric mode resonances and lack of magnetic modes[28] enable single-mode selection over a relatively broad wavelength range of 450-650 nm, thereby improving signal detection (gray region in the $s$-polarized CL spectrum at $\theta$ = 20° with sphere-edge excitation; Figure 3(b)). The corresponding CL photon map exhibits high contrast at the sphere edges along the $y$-direction, indicating excitation of an in-plane electric dipole oriented along the $y$-axis (Figure 3(d)). The high-contrast regions extend slightly beyond the sphere boundary (dotted black curve), reflecting the nature of the electric dipole, whose oscillating electric fields extend partially outside the particle[29]. The slight distortion of the dotted curve results from scanning drift during data acquisition. In addition to the $y$-dipole contrast, some intensity is also visible at the bottom of the sphere, corresponding to the excitation of the $x$-oriented electric dipole of which $s$-polarized component is included at $\phi \neq 0°$ angles[12]. For phase retrieval, we utilize this $x$-dipole since it has opposite $s$-polarized field phases at $\phi > 0°$ and $\phi < 0°$ angles.

The emission and angular images used for phase retrieval (Figure 3(e, f)) were acquired under bottom-edge excitation (red dot in Figure 3(c)) within the wavelength range highlighted in Figure 3(b). The measured intensity distributions in both spaces agree well with simulations calculated at $\lambda$ = 530 nm, exhibiting double emission spots in real space (Figure 3(e, h)) and azimuthal-side emission in angular space (Figure 3(f, i)). The emission spots are slightly tilted along the $y'$-direction due to an unintended azimuthal misalignment of the pinhole mask along the $Y$-axis, placing the detection position approximately at ($\theta$, $\phi$) = (20°, –35°), as shown in Figure 3(f, i). Consequently, the angular emission distribution is also slightly shifted from the center.

The phase distribution retrieved from the intensity data in Figure 3(e, f) is presented in Figure 3(g) and agrees well with the simulated phase distribution in Figure 3(j), with the off-center pinhole position taken into account. The retrieved phase exhibits a discontinuity across the azimuthal direction at $\phi = 0°$. Quantitative line profiles extracted along the arrows in Figure 3(g, j) further reveal a relative phase jump of approximately $\Delta\psi \approx \pi$ rad (Figure 3(k)). The retrieved phase profile (green curve) displays a more gradual transition compared to the sharp discontinuity observed in the simulated profile (orange curve) due to noise in the angular intensity image (Figure 3(f)); see the SI for an evaluation of the noise effect. This phase jump originates from the intrinsic symmetry of the dipole radiation pattern. In the vicinity of the nodal direction (near $\phi = 0°$ in Figure 3(f, i)), the emission amplitude approaches zero, and the electromagnetic phase becomes undefined. As the field crosses this node, its sign reverses, resulting in a relative π phase discontinuity across the azimuthal direction (Figure 3(g, j))[30]. These results indicate that the optical CL wavefront of dipole radiation exhibits an out-of-phase inversion and further demonstrate the robustness of the iterative phase-retrieval algorithm, which accurately reconstructs phase distributions despite slight experimental misalignment.

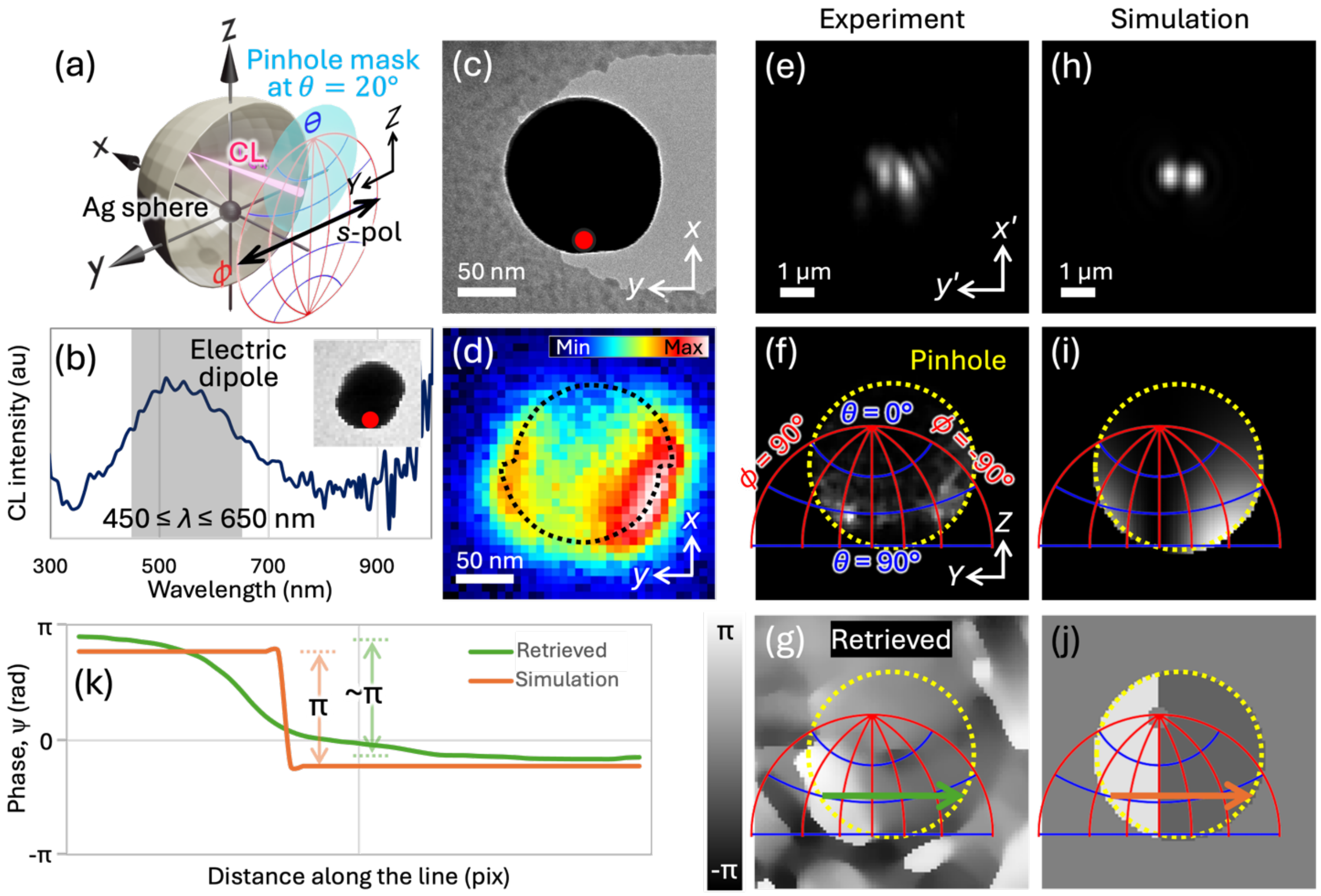

**Figure 3**. Electric dipole from a silver sphere. (**a**) Schematic illustration of a silver sphere inside a parabolic mirror. The blue circle indicates a pinhole mask with a detection solid angle of 2.4 sr at ($\theta$, $\phi$) = (20°, 0°), and the black arrow indicates an *s*-polarizer aligned along the *y*-axis. (**b**) Corresponding CL spectrum acquired with sphere-edge excitation, as shown in the inset, at $\theta = 20°$. The gray rectangle indicates the selected wavelength range of 450 nm ≤ λ ≤ 650 nm. (**c**) STEM bright-field image of a silver sphere with a diameter of 140 nm. The red dot indicates the electron beam position. (**d**) CL photon map integrated over the wavelength range indicated in (b) at $\theta = 20°$. Experimental (**e**) emission image and (**f**) angular image acquired at the beam position shown in (c) within the wavelength range indicated in (b), along with the corresponding (**g**) retrieved phase image. Simulated (**h**) emission image, (**i**) angular image, and (**j**) phase image. (**k**) Line profiles of the phase distribution extracted along the directions of the arrows in (g, j). The angular-space images in (f, g,

i, j) also show the off-centered pinhole mask (yellow circle) positioned at ($\theta$, $\phi$) = (20°, -35°), along with the upper half of the projected angular coordinates. The images in (e-j) are magnified by approximately ×3.1 relative to the original size used in the phase-retrieval process.

To assess the phase distribution of optical CL wavefronts emitted by an antisymmetric plasmon mode (i.e., a standing SPP mode with antinodes at the grating step edges), a one-dimensional (1D) silver plasmonic crystal was investigated. The structure has a periodicity of $P$ = 600 nm, height $H$ = 100 nm, and terrace width $D$ = 420 nm (Figure 4(c)). A ~200 nm thick silver film was vapor deposited onto the structure to support SPPs and capped with a 5 nm $SiO_2$ layer to prevent surface degradation (backscattered electron image in Figure 4(c)). The grating terraces were oriented perpendicular to the parabolic mirror axis along the $x$-direction; accordingly, a $p$-polarizer was employed to select the $xz$-component of the electric field[9] (Figure 4(a)). The emission direction was defined using a pinhole mask positioned at ($\theta$, $\phi$) = (20°, 0°).

The CL spectrum measured at $\theta$ = 20°, integrated over the entire scanning area, exhibits two peaks arising from band splitting at the Γ point ($\theta$ = 0°) of the dispersion relation[31] (Figure 4(b)). Although the measurements were performed at $\theta$ = 20° to capture a large fraction of the angular emission pattern, we note that this angle remains sufficiently close to the Γ point for the resonant plasmon modes to retain predominantly standing-wave characteristics[14]. The antisymmetric (A) mode was selected for phase analysis because of its relatively simple charge distribution, in which a series of in-plane electric dipoles is excited parallel to the $x$-axis. This dipolar excitation is evidenced by the bright contrasts at the terrace edges in the CL photon map (Figure 4(d)), integrated over 650-800 nm at $\theta$ = 20° (gray region in Figure 4(b)).

Based on the A-mode charge distribution, the electron beam was positioned at the terrace-front edge (red dot in Figure 4(c)) to acquire the emission and angular intensity images (Figure 4(e, f)) for phase retrieval[14]. The measured results agree well with the corresponding simulations (Figure 4(h, i)). The simulations modeled five in-plane electric dipoles oriented along the $x$-axis and separated by the grating periodicity ($P$ = 600 nm) at a photon wavelength 712 nm. Each dipole was assigned a relative phase of $\pm k_{SPP}P$, representing decoupled propagating SPPs with wavevector $k_{SPP}$[14]. The emission intensity was assumed to decay radially from the excitation point, consistent with spherical propagation of SPP-mediated radiation, while attenuation losses were neglected because the SPP propagation length on silver is sufficiently longer than the measured sample size[32]. Strong coherent interference with TR, modeled as an out-of-plane electric dipole oriented along the $z$-axis at the excitation point, was also included. Consequently, two predominant emission spots are observed in real space (Figure 4(h)), with most of the radiation directed near $\theta = 0°$ in angular space (Figure 4(i)). The slight differences in the angular images arise from sample-angle misalignment and the use of a wide detection bandwidth in the experiment, whereas the calculations were performed at a single wavelength (see the SI for details).

The retrieved phase distribution is shown in Figure 4(g). A quantitative comparison with the simulated phase (Figure 4(j)) along the horizontal solid green and dotted orange arrows, where the angular radiation is predominantly concentrated, reveals a nearly homogeneous phase profile with a relative phase difference of approximately $\Delta\psi \approx 0$ rad (Figure 4(k)). The slight dip in the retrieved phase profile (arrow #1) arises from the electron-beam entrance hole in the parabolic mirror. Additionally, the simulated angular image exhibits weak radiation near $\theta = 90°$ (Figure 4(i)), which is also present in the measured angular image (Figure 4(f); see SI for a clearer intensity distribution). Phase line profiles extracted along the vertical solid pink and dotted black arrows indicate a relative phase shift of less than π rad between the upward emission at $\theta = 0°$ (around arrow #2) and the in-plane emission near $\theta = 90°$ (around arrow #3). These results demonstrate that the optical CL wavefronts of the A-mode exhibit a nearly constant relative phase along the azimuthal ($\phi$) direction while varying along the polar ($\theta$) direction.

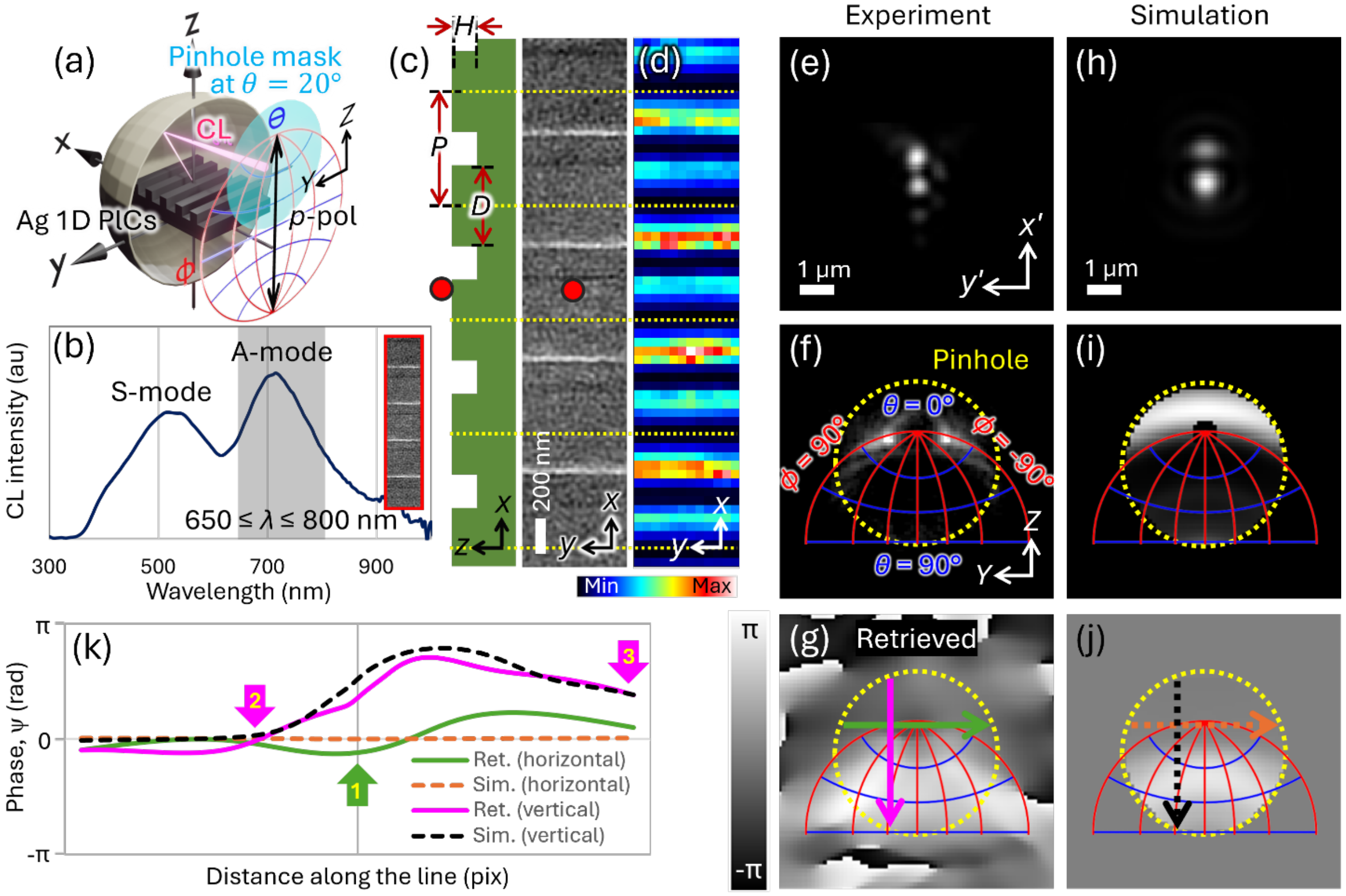

**Figure 4**. An antisymmetric mode excited on a 1D silver plasmonic crystal. (**a**) Schematic illustration of the 1D silver plasmonic crystal, with the grating terraces oriented perpendicular to the parabolic mirror axis along the *x*-direction. The blue circle indicates the pinhole mask defining a detection solid angle of 2.4 sr at $(\theta, \phi) = (20°, 0°)$, and the black arrow denotes a *p*-polarizer aligned in the *xz*-plane. (**b**) Area-integrated CL spectrum acquired at $\theta = 20°$, exhibiting symmetric (S) and antisymmetric (A) plasmonic band-edges modes. The gray rectangle marks the selected wavelength range of 650 nm ≤ λ ≤ 800 nm. (**c**) Structural illustration and backscattered electron image of the 1D silver plasmonic crystal with periodicity $P = 600$ nm, height $H = 100$ nm, and terrace width $D =$

420 nm. The dotted yellow lines indicate the terrace centers, and the red dot marks the electron-beam position. (**d**) CL photon map integrated over the wavelength range indicated in (b) at $\theta = 20°$. Experimental (**e**) emission image and (**f**) angular image acquired at the beam position shown in (c) within the wavelength range indicated in (b), together with (**g**) the corresponding retrieved phase image. Simulated (**h**) emission image, (**i**) angular image, and (**j**) phase image. (**k**) Line profiles of the phase distribution extracted along the horizontal and vertical arrows indicated in (g, j). The angular-space images in (f, g, i, j) also display the pinhole mask (yellow circle) and the upper half of the projected angular coordinates. The images in (e-j) are magnified by approximately ×3.1 relative to the original size used in the phase-retrieval process.

To further analyze more complex phase distributions, a surface plasmon mode in a silver nanowire was investigated. The nanowire (commercial product A50SL from Filgen, dispersed in ethanol) has a length of 1410 nm and a diameter of 130 nm (STEM bright-field image in Figure 5(c)). We employed *s*-polarization parallel to the wire's long axis (along the *y*-axis) to extract the propagating modes (Figure 5(a)). Plasmon resonances with induced charge oscillations longitudinal to the wire axis are excited for both odd and even mode orders ($m$) as presented in the *s*-polarized wavelength-scan line profile measured at $\theta = 20°$ (Figure 5(b)), where resonances from $m = 1$ to 6 are observed. The higher-order $m = 4$ mode within the wavelength bandwidth of 450-550 nm was selected for this study (gray region in Figure 5(b); CL spectrum in the SI). The corresponding CL photon map (Figure 5(d)) exhibits two hotspots at the end facets, attributed to antinode end bunching[33], and three additional hotspots along the inner wire body, attributed to surface plasmon standing waves[8] (wire body is indicated by the dotted black curve).

The emission and angular images used for phase retrieval (Figure 5(e, f)) were acquired under wire-center excitation (red dot in Figure 5(c)), within the wavelength range highlighted in Figure 5(b), corresponding to an antinode of the $m = 4$ standing wave. The simulation was performed at a photon wavelength 500 nm using two electric dipoles located at the end facets, each positioned 705 nm from the wire center and oriented along the wire's long axis[4, 34]. The dipoles were assigned equal amplitudes because the wire-center-excited SPPs travel equal distances to the two end facets and have an opposite relative phase of 180° for even-order modes. Good agreement is observed between the experimental data and the simulation, showing double emission spots in real space (Figure 5(e, h)) and four lobes in angular space with destructive interference at $\phi = 0°$ (Figure 5(f, i)); see also the SI for their intensity line profiles. The discrepancy in emission intensities can be attributed to the non-identical, irregular morphology of the nanowire end facets, which modify the SPP reflection efficiency and results in unequal dipole radiation strengths at the wire tips, leading to asymmetric emission[15]. The inclusion of a wide detection bandwidth is also discussed in the SI.

The phase distribution retrieved from the intensity data in Figure 5(e, f) is presented in Figure 5(g) and agrees well with the simulated phase distribution in Figure 5(j), exhibiting three discontinuity features across the azimuthal direction $\phi$. Line profiles extracted along the arrows in Figure 5(g, j) further reveal the three expected relative phase jumps of $\Delta\psi \approx \pi$ rad (Figure 5(k)), which occur because the emission amplitude approaches zero near the nodal directions, causing a sign reversal of the field and a π phase shift, as discussed previously. These results indicate that the CL radiation associated with coherently interfering nanowire dipoles in the far field exhibits out-of-phase wavefronts between neighboring angular emission lobes, further demonstrating the capability of the CL-accommodated algorithm to retrieve complex phase distributions.

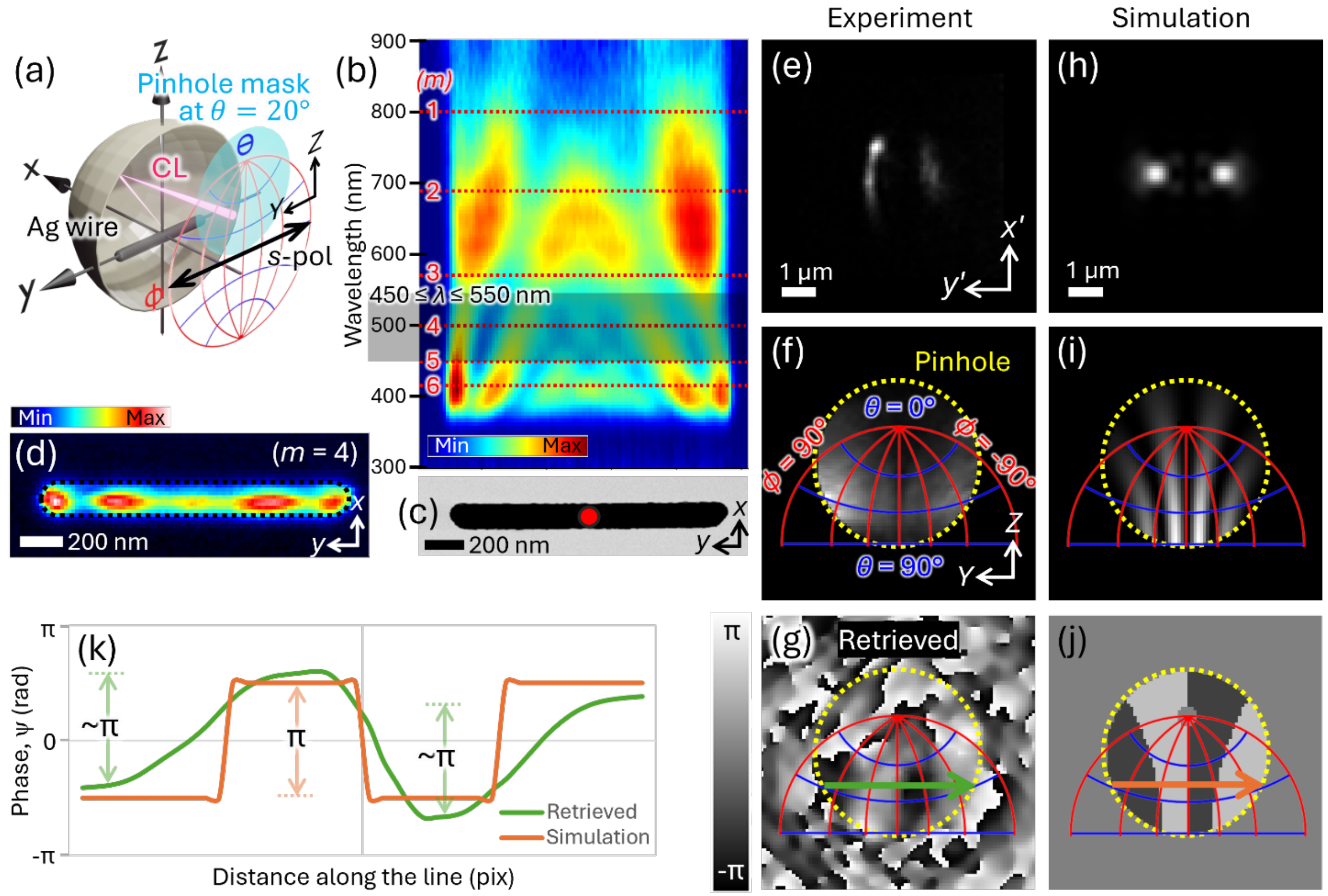

**Figure 5**. Surface plasmon mode from a silver nanowire. (**a**) Schematic illustration of the nanowire inside a parabolic mirror. The emitted CL is selected by a pinhole mask with a detection solid angle of 2.4 sr located at ($\theta$, $\phi$) = (20°, 0°), indicated by the blue circle, and filtered by an *s*-polarizer (black arrow) to select light polarized parallel to the wire body along the *y*-axis. (**b**) Polarized CL wavelength-scan line profile measured at $\theta$ = 20°, integrated over the entire nanowire body shown in the STEM bright-field image in (**c**), which depicts a silver nanowire with a length of 1410 nm and a diameter of 130 nm. The gray rectangle in (b) indicates the selected wavelength range of 450 nm ≤ λ ≤ 550 nm, isolating the plasmon mode of order $m$ = 4. (**d**) CL photon map measured at $\theta$ = 20°, integrated over the wavelength range indicated in (b). Experimental (**e**) emission image and (**f**) angular image acquired at the beam position shown in (c) within the wavelength range indicated in (b), together with the corresponding (**g**) retrieved phase image. Simulated (**h**) emission image, (**i**)

angular image, and (**j**) phase image. (**k**) Line profiles of the phase distribution extracted along the directions indicated by the arrows in (g, j). The angular-space images in (f, g, i, j) also show the pinhole mask (yellow circle) and the upper half of the projected angular coordinates. The images in (e-j) are magnified by approximately ×3.1 relative to the original size used in the phase-retrieval process.

In conclusion, we have demonstrated a reference-free phase retrieval approach for CL wavefronts based on the error-reduction Gerchberg-Saxton algorithm implemented with experimentally measured CL data in a STEM. Using a dataset consisting of real-space and angular-space intensity images, the phase of emissions associated with electric dipoles in various nanostructures, including a planar surface, nanosphere, plasmonic crystal, and nanowire, can be directly retrieved. The reconstructed phase distributions reveal distinct characteristics, such as a uniform relative phase, a number of $\pi$ phase jumps, and a varying phase shift across the ($\theta$, $\phi$) space, depending on the underlying radiation mechanisms. This work establishes a robust and flexible phase retrieval framework that does not rely on a reference wave or interferometric measurements, making it readily applicable for reconstructing the phase of optical CL fields in a wide range of electron-beam-excited nanophotonic structures. Furthermore, by enabling direct access to the phase of CL emission, the approach expands the capabilities of CL spectroscopy and provides a powerful tool for studying optical CL wavefront structures and radiation mechanisms, which may further support developments in nanoscale wavefront engineering.

## AUTHOR INFORMATION

**Corresponding authors**

**Izzah Machfuudzoh**

Department of Materials Science and Engineering, School of Materials and Chemical Technology, Institute of Science Tokyo, Yokohama 226-8503 Japan

Email: machfuudzoh.i.aa@m.titech.ac.jp

ORCID: https://orcid.org/0009-0000-8761-0036

**Takumi Sannomiya**

Department of Materials Science and Engineering, School of Materials and Chemical Technology, Institute of Science Tokyo, Yokohama 226-8503 Japan

Email: sannomiya@mct.isct.ac.jp

ORCID: https://orcid.org/0000-0001-7079-2937

**Author**

**Ryoichi Horisaki**

Department of Information Physics and Computing, Graduate School of Information Science and Technology, The University of Tokyo, Tokyo 113-8656 Japan

ORCID: https://orcid.org/0000-0002-2280-5921

## ASSOCIATED CONTENT

**Supporting information**

Performance evaluation of the phase-retrieval algorithm (including the uniqueness of the solution, phase ramps in angular space, and robustness to noisy angular images), as well as further information on the plasmonic crystal (including the angular intensity profile and wavelength-integrated simulations) and the nanowire (including the spectrum, intensity line profiles, and wavelength-integrated simulations).

**Competing interests**

The authors declare no financial conflicts of interest.

## Funding

This work is financially supported by JST CREST (JPMJCR25I3), JST FOREST (JPMJFR213J, JPMJFR2448), and JSPS Kakenhi (23K26567, 24H00400).

## Author contributions

T.S. and R.H. conceived the project. T.S. supervised the project and developed the simulation code. I.M. performed the experiments and developed the phase-retrieval algorithm code. T.S. and I.M. analyzed the data. I.M. wrote the original manuscript, and all authors discussed the results and contributed to the manuscript.

# TOC Graphic

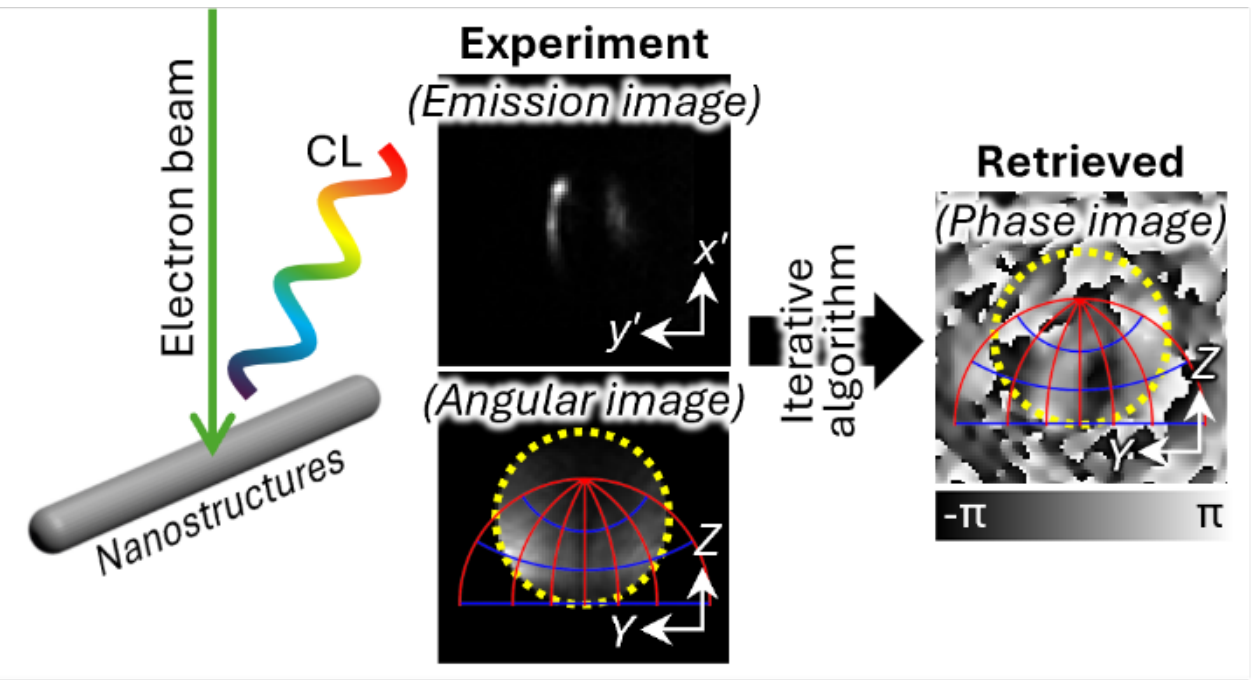
Electron beam
CL
Nanostructures
Experiment
(Emission image)
x′
y′
(Angular image)
Z
Y
Iterative algorithm
Retrieved
(Phase image)
Z
Y
-π
π

# Supporting Information

# Cathodoluminescence Wavefront Retrieval

*Izzah Machfuudzoh[1,]*, Ryoichi Horisaki[2], Takumi Sannomiya[1,]***

[1] Department of Materials Science and Engineering, School of Materials and Chemical Technology, Institute of Science Tokyo, 4259 Nagatsuta, Midori-ku, Yokohama 226-8503 Japan

[2] Department of Information Physics and Computing, Graduate School of Information Science and Technology, The University of Tokyo, 7-3-1 Hongo, Bunkyo-ku, Tokyo 113-8656 Japan

**Corresponding authors**

* Izzah Machfuudzoh (Email: machfuudzoh.i.aa@m.titech.ac.jp)

** Takumi Sannomiya (Email: sannomiya@mct.isct.ac.jp)

## The uniqueness of the solution

The uniqueness of the phase-retrieval solution was evaluated by initializing the algorithm with different random phase distributions, each within the range of [-π, π]. To facilitate the evaluation, a simulated dataset exhibiting an abrupt phase variation was employed. For this purpose, an in-plane electric dipole with *s*-polarization at a photon wavelength of $\lambda = 600$ nm was used as the emission source in the simulation. A pinhole mask configuration identical to that used in the experimental setup (Figure 1(a-c)), corresponding to a detection solid angle of 2.4 sr positioned at an emission angle of $(\theta, \phi) = (20°, 0°)$, was incorporated into the simulation and is indicated by yellow circles in Figure S1(b-h).

The simulated intensity distributions of the emission and angular images are shown in Figure S1(a, b), respectively. Their corresponding amplitudes were used as inputs for the phase-retrieval algorithm, yielding multiple reconstructed phase images obtained from different initial random phase conditions (Figure S1(d-h)). The reconstructed phase images exhibit an identical vertical central-line feature consistent with the calculated phase distribution in Figure S1(c). To quantitatively evaluate the phase distributions, line profiles were extracted along the directions indicated by arrows in Figure S1(c-h). All reconstructed phase images exhibit a phase-jump feature with a relative phase difference of $\Delta\psi = \pi$ rad across the azimuthal angle at $\phi = 0°$, as shown in Figure S1(i), consistent with the calculated phase distribution. These results confirm that the relative phase structure is accurately reconstructed. Note also that the retrieved phase distributions within the pinhole mask differ only by a global phase offset while preserving the same relative phase (Figure S1(i)). This behavior presents the well-known uniqueness ambiguity in phase retrieval[1-3], which arises because intensity measurements are insensitive to an overall phase factor of the optical field and therefore remain invariant under a global phase shift. Consequently, physically meaningful information is contained in the relative phase distribution rather than in the absolute phase.

The convergence behavior of the algorithm was further evaluated by calculating the MSE between the amplitudes derived from the simulated emission and angular intensity distributions with the corresponding reconstructed amplitudes in real and angular spaces, respectively (Figure S1(j, k)). Both domains exhibit small residual errors on the order of $10^{-12}$ after 100 iterations, confirming that the algorithm converges to a stable solution and that the reconstructed complex optical field, encompassing both amplitude and phase, is consistent with the measured data (i.e., the simulation).

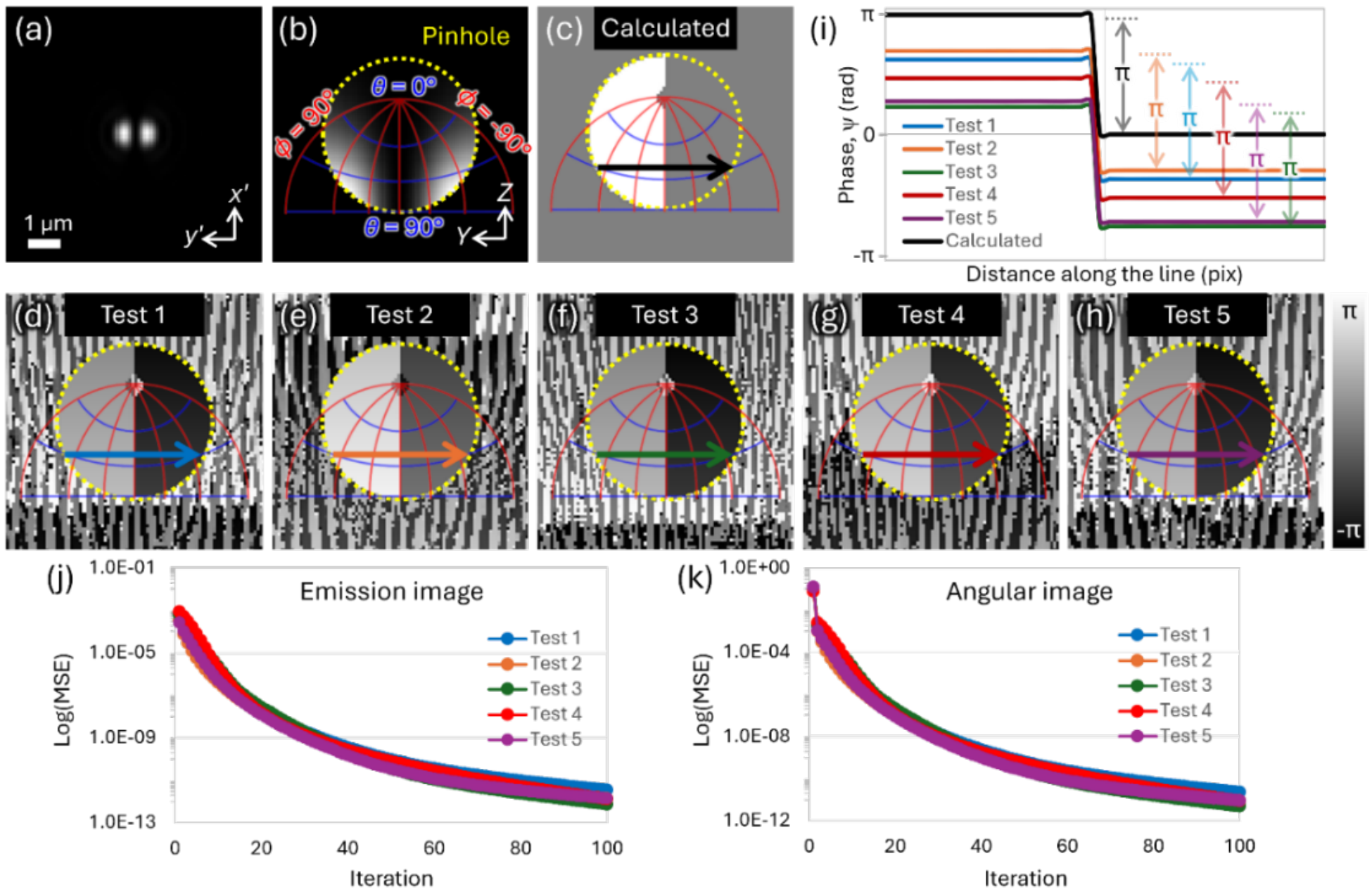

**Figure S1**. Simulated (**a**) emission image, (**b**) angular image and (**c**) phase distribution. (**d-h**) Phase distributions retrieved by the algorithm using different initial random phase conditions. (**i**) Line profiles of the phase distributions extracted along the directions indicated by arrows in (c-h), within the pinhole mask region marked by yellow circles. Convergence analysis showing the logarithm of the mean-squared error (MSE) as a function of iteration number for (**j**) emission images and (**k**) angular images. Colors in (i-k) indicate the corresponding phase-retrieval tests shown in (d-h). The upper half of the projected angular coordinates are overlaid in (b-h) for reference. The images are magnified by approximately ×3.1 relative to the original size used in the phase-retrieval process.

## Phase ramps in angular space

The Fourier shift theorem states that a spatial shift in real space induces a linear phase modulation in its Fourier-conjugate space[4]. In the present context, this means that if the emission spot(s) are displaced from the origin in the *x'y'z'* emission coordinates (real space), the corresponding phase distribution in angular space will exhibit linear phase ramps. To understand this effect, we consider the same simulation parameters presented in Figure S1, consisting of an *s*-polarized in-plane electric dipole at $\lambda$ = 600 nm. When the emission spots are located at (*x'*, *y'*, *z'*) = (0, 0, 0) μm, the retrieved phase shows a homogeneous distribution within each region separated by the intrinsic phase jump (Figure S2(a)). However, when the emission spots are shifted along the *x'*-direction to (*x'*, *y'*, *z'*) =

(1.4, 0, 0) μm, the retrieved phase in angular space exhibits additional linear fringes within the pinhole mask (yellow circle), aligned with the direction of the spatial shift (Figure S2(b)). Similarly, a shift along the *y'*-direction to (0, –1.4, 0) μm (Figure S2(c)), as well as diagonal shifts to (1.4, –1.4, 0) μm (Figure S2(d)) and (–1.4, –1.4, 0) μm (Figure S2(e)), introduces linear phase ramps whose orientation and gradient directly reflect the direction and magnitude of the spatial displacement. Although phase ramps do not inherently alter the underlying physical phase relationships, minimizing spatial shifts in real space simplifies the interpretation of the retrieved phase distributions in angular space.

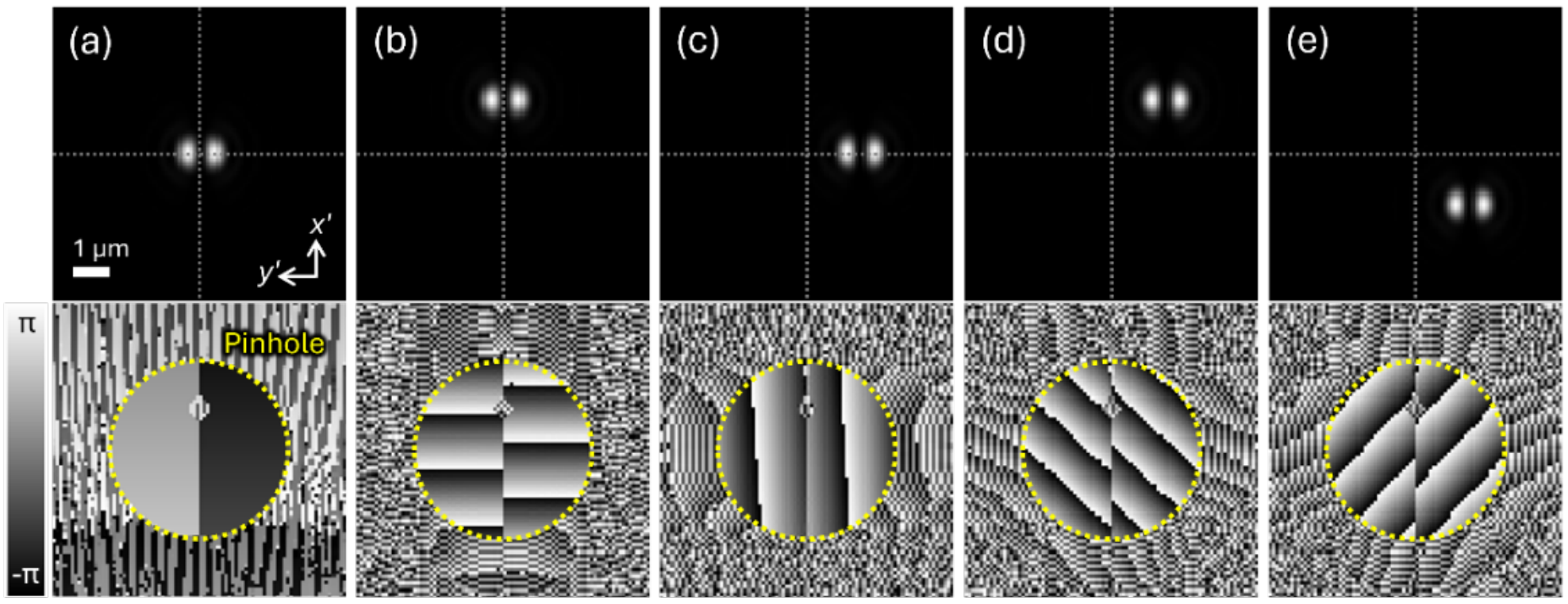

**Figure S2.** Simulated emission images (top row) and the corresponding retrieved phase images (bottom row) for (**a**) non-shifted emission spots and for emission spots shifted by 1.4 μm in the (**b**) positive *x'*- and (**c**) negative *y'*-directions, as well as (**d**) northeast and (**e**) southeast directions. The yellow circles indicate the region defined by the pinhole mask with a detection solid angle of 2.4 sr positioned at ($\theta$, $\phi$) = (20°, 0°). All images are magnified by approximately ×3.1 relative to the original size used in the phase-retrieval process.

## Robustness to noisy angular images

Parseval's theorem requires conservation of the total optical intensity between two Fourier-conjugate domains, such that the same photon budget used to record emission spots in real space is distributed over a larger angular emission pattern in angular space. This redistribution often leads to reduced photon statistics per pixel in angular images, making them more susceptible to noise. Hence, it is essential to assess the algorithm's performance under noisy angular imaging conditions.

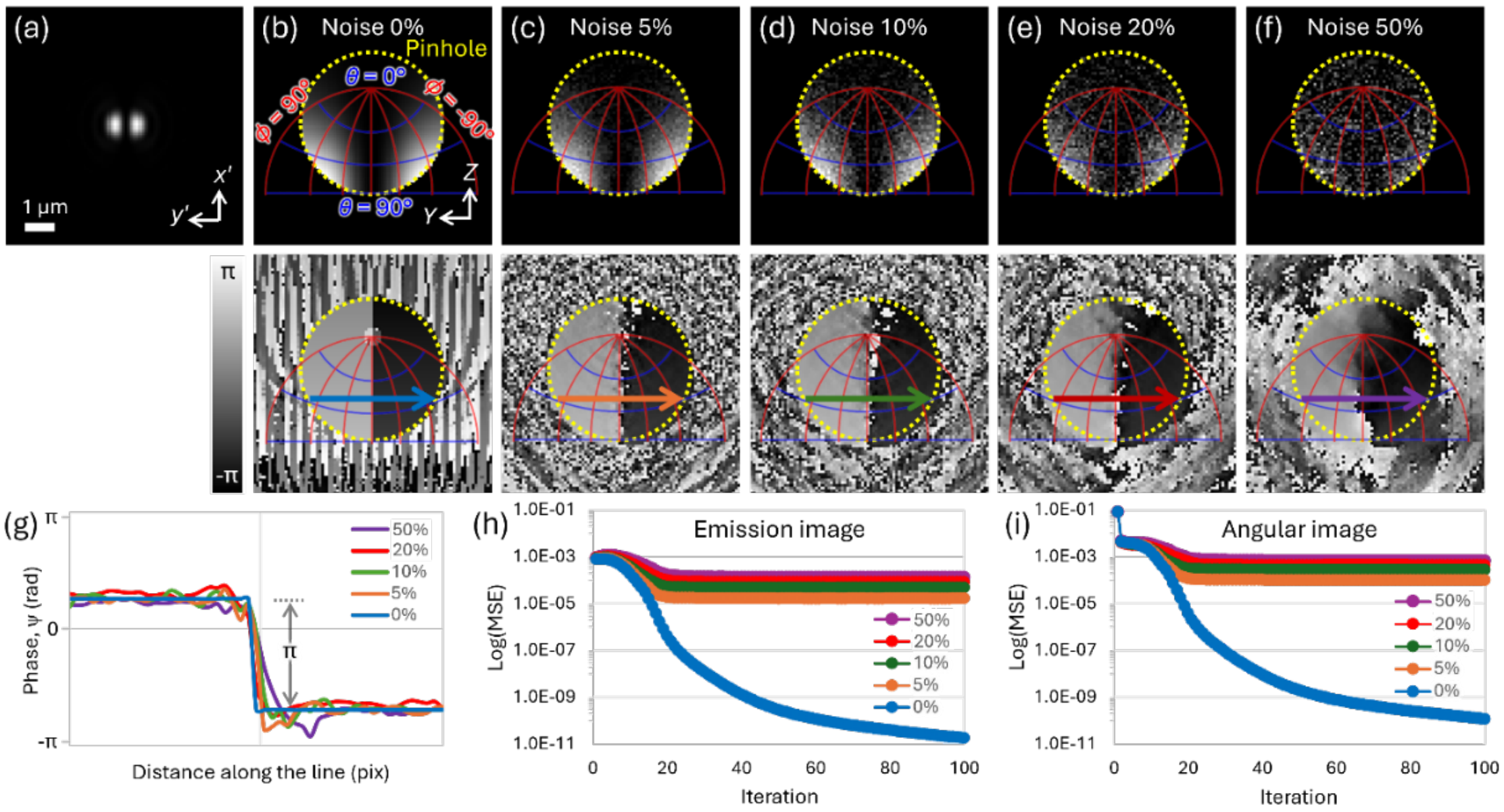

**Figure S3**. (**a**) Simulated emission image in real space, and reconstructed phase distributions in angular space (bottom row) obtained using angular images (top row) with Gaussian noise levels of (**b**) 0%, (**c**) 5%, (**d**) 10%, (**e**) 20%, and (**f**) 50%. Noise was added only within the pinhole mask region marked by yellow circles. (**g**) Line profiles of the phase distributions extracted along the directions indicated by arrows in (b-f). Convergence analysis for (**h**) emission images and (**i**) angular images, showing the logarithm of the mean-squared error (MSE) between the simulated and reconstructed amplitudes as a function of iteration number. Colors in (g-i) correspond to the phase-retrieval cases with different noise levels shown in (b-f). The upper half of the projected angular coordinates are overlaid in (b-f) for reference. The images are magnified by approximately ×3.1 relative to the original size used in the retrieval process.

Gaussian noise was introduced into the simulated angular intensity images as a simplified model of stochastic measurement uncertainty[5]. The noise was applied only within the pinhole mask region, indicated by yellow circles. The noise level was controlled by scaling the standard deviation of the noise relative to the root-mean-square (RMS) value of the noise-free angular intensity distribution shown in the top image of Figure S3(b). This procedure generated angular intensity images with noise levels ranging from 5% to 50%, as shown in the top row of Figure S3(c-f).

Using the amplitude of the noise-free emission intensity image in Figure S3(a) together with a fixed initial random phase as the real-space input, the phase distributions corresponding to each angular imaging condition were retrieved (bottom row of Figure S3(b-f)). Line profiles extracted along the arrow directions confirm that the relative phase jump of $\Delta\psi = \pi$ rad across the azimuthal angle at $\phi = 0°$ is preserved in all cases (Figure S3(g)). However, the phase distributions retrieved with noisy angular images (Figure S3(c-f)) exhibit reduced spatial homogeneity, as evidenced by increased fluctuations in the phase profiles (Figure S3(g)). Furthermore, the phase-jump transition becomes progressively smoother and the phase distributions extend slightly beyond the pinhole mask region with increasing noise level. Nevertheless, the retrieved phase distributions consistently preserve the expected phase-jump feature, demonstrating that the relative phase structure remains recoverable despite noise contamination in the angular intensity images.

The convergence behavior of the algorithm was further evaluated, as presented in Figure S3(h) for the emission images and Figure S3(i) for the angular images. The reconstruction errors obtained using noisy angular images (curves shown in colors other than blue in Figure S3(h, i)) initially decrease and then saturate at levels approximately six orders of magnitude higher than those of the noise-free case (blue curves) after 100 iterations. This behavior arises because the Fourier modulus (the noisy angular images in Figure S3(c-f)) is inconsistent with its counterpart in the object-domain constraints (the noise-free emission image in Figure S3(a)) due to the presence of noise[6]. Despite this increase, the algorithm still converges to relatively low residual errors on the order of $10^{-5}$, demonstrating the robustness of the phase-retrieval process against noise in angular images.

## Angular intensity profile and wavelength-integrated simulations of the plasmonic crystal

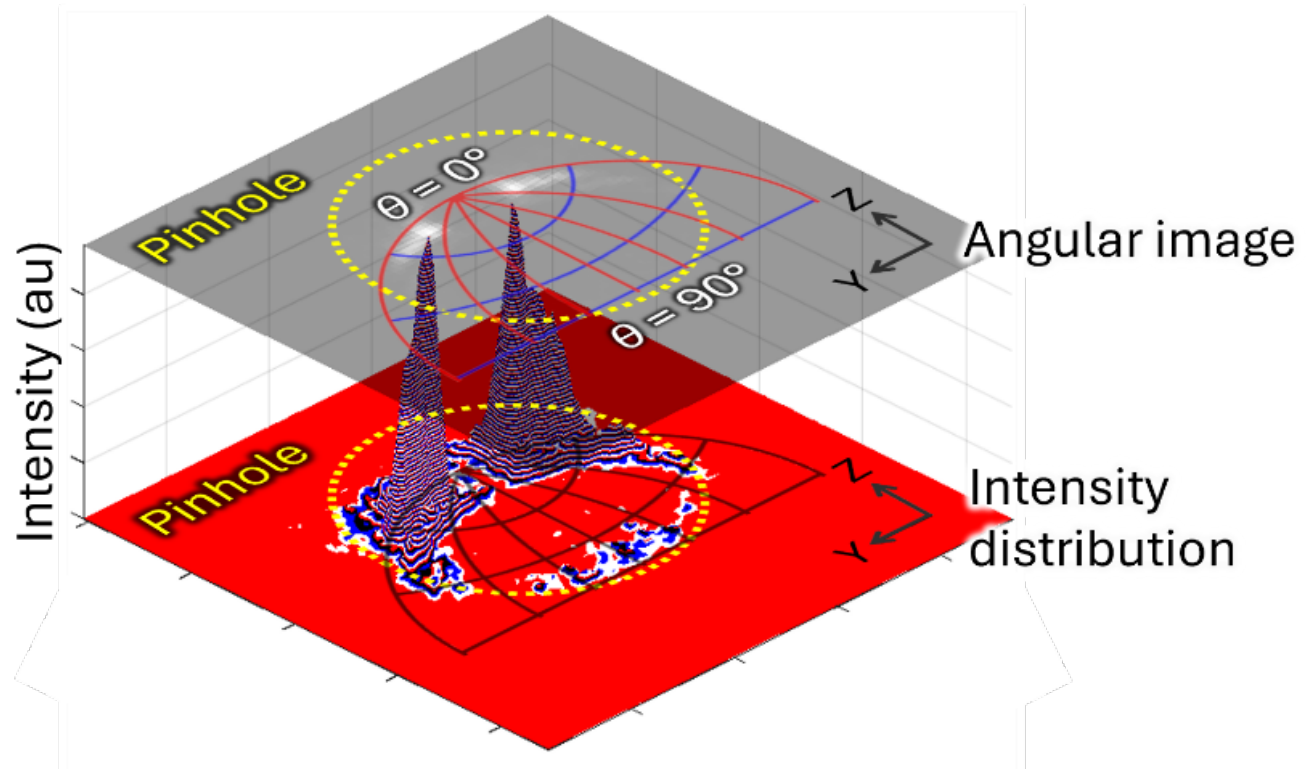

**Figure S4.** Intensity profile (bottom panel) extracted from the angular image (top panel) of the emission from the 1D silver plasmonic crystal. The upper half of the projected angular coordinates and the pinhole mask indicated by the yellow circle are also shown.

The intensity distribution of the experimental angular image acquired under electron-beam excitation at the terrace-front edge of the 1D silver plasmonic crystal (Figure 4(f) in the main text) is shown in Figure S4. In addition to the strong intensity observed at upward emission angles around $\theta = 0°$, measurable intensity is also present near the lateral angles around $\theta = 90°$, in agreement with the simulation shown in Figure 4(i). Accordingly, the relative phase analysis across the polar $\theta$ direction (i.e., the line profiles of the phase distributions in Figure 4(k), extracted along the vertical arrows in Figure 4(g, j)) is physically meaningful.

The emission, angular, and phase images from the same 1D silver plasmonic crystal, simulated over the wavelength range 600 nm $\leq \lambda \leq$ 750 nm, are shown in Figure S5. While both the emission and phase images (Figure S5(a, c)) maintain intensity and phase distributions similar to those calculated at the resonance wavelength (Figure 4(h, j) in the main text), the angular image in Figure S5(b) shows a slight difference compared with Figure 4(i), with more emission concentrated at $\theta = 0°$ and along $\phi = \pm 90°$. This discrepancy is attributed to the inclusion of the wavelength bandwidth in the calculation, consistent with the use of bandpass filters in the experiment, resulting in very good agreement between the measured angular image (Figure 4(f)) and the simulated wavelength-integrated image (Figure S5(b)). Note that the wavelength range used in the simulation was shifted slightly toward shorter wavelengths by 50 nm compared with that used in the experiment (650 nm $\leq \lambda \leq$ 800 nm); nevertheless, the A-mode is still included (see the CL spectrum in Figure 4(b)).

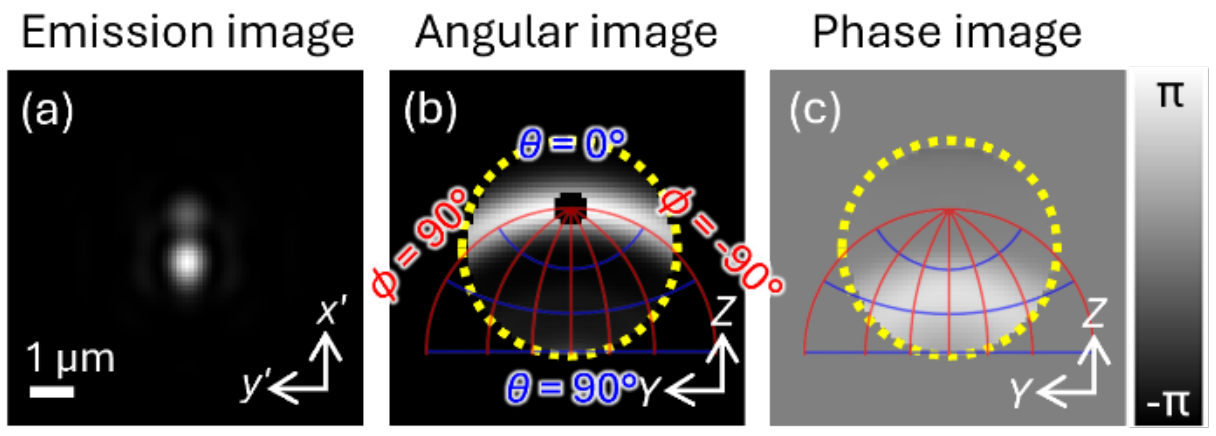

**Figure S5.** Wavelength-integrated (**a**) emission image, (**b**) angular image, and (**c**) phase image of the 1D silver plasmonic crystal, calculated over the range 600 nm $\leq \lambda \leq$ 750 nm. The yellow circle indicates the pinhole mask.

## Spectrum, intensity line profiles, and wavelength-integrated simulations of the nanowire

The *s*-polarized CL spectrum of the silver nanowire with a length of 1410 nm and a diameter of 130 nm, presented in Figure 5(c) of the main text, is shown in Figure S6. The spectrum, measured at an emission angle of $(\theta, \phi) = (20°, 0°)$ using a pinhole mask with a detection solid angle of 2.4 sr, is integrated over the entire wire as indicated in the inset. The wavelength bandwidth of 450 nm ≤ $\lambda$ ≤ 550 nm used in the experiment to isolate the resonant $m = 4$ mode is indicated in the spectrum by a gray rectangle.

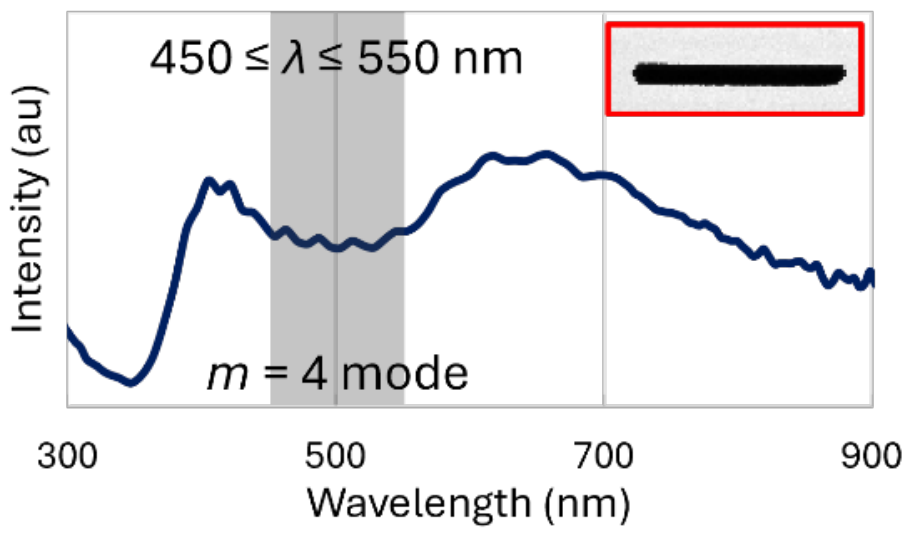

**Figure S6.** Area-integrated, *s*-polarized CL spectrum of a silver nanowire measured at $(\theta, \phi) = (20°, 0°)$. The wavelength bandwidth used to isolate the $m = 4$ surface plasmon mode is indicated in gray color.

The intensity line profiles of the emission and angular images from the same silver nanowire, presented in Figure 5(e, h) and Figure 5(f, i) of the main text, are shown in Figure S7(a, b) and Figure S7(c, d), respectively, for both the experimentally acquired data and the corresponding simulations. The measured emission image (Figure S7(a)) shows two well-separated intensity peaks (i.e., emission spots) in real space that agree well with the simulation (Figure S7(b)). The angular images also show similarly good agreement, exhibiting four intensity peaks (i.e., emission lobes) across the azimuthal direction in angular $(\theta, \phi)$ space (Figure S7(c, d)). The intensity line profiles were extracted along the directions indicated by the respective green and orange arrows.

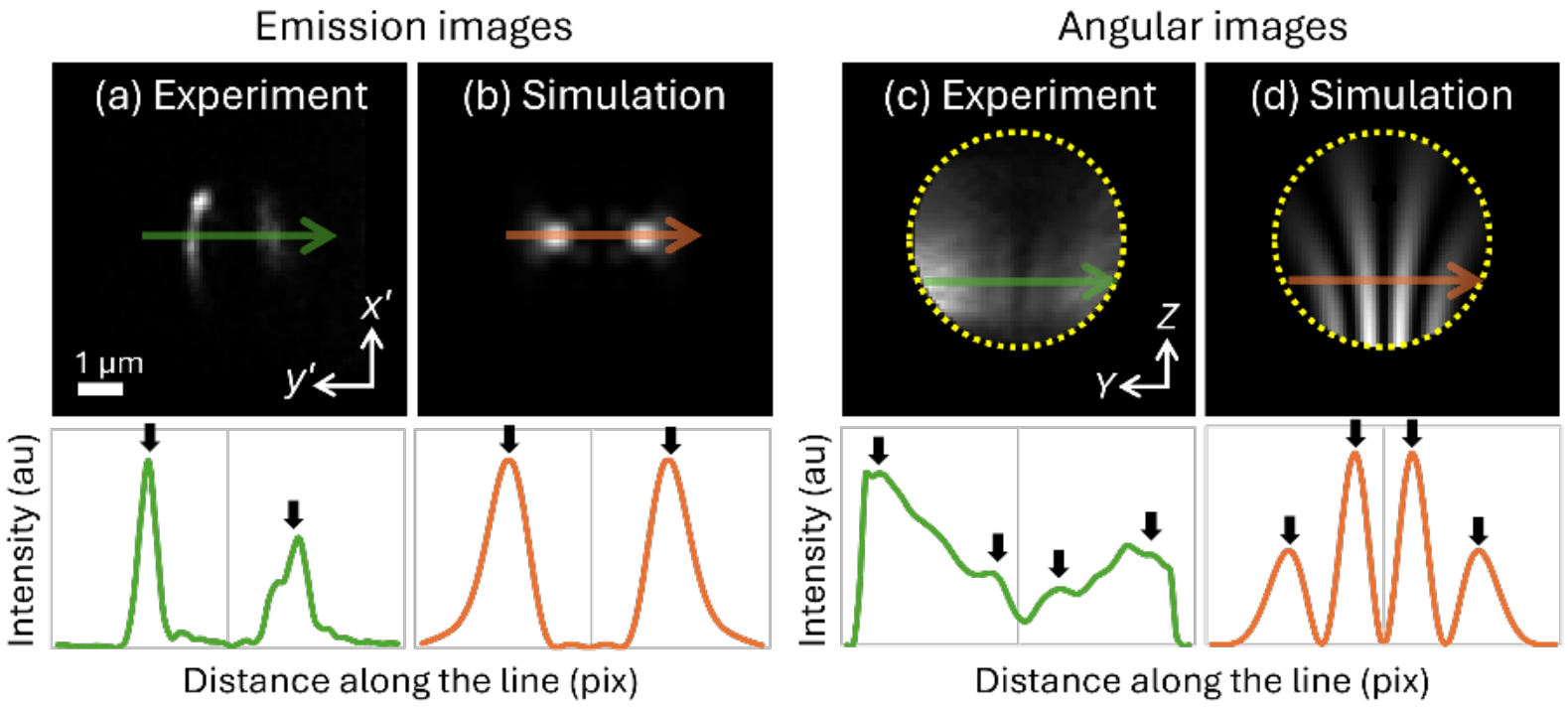

**Figure S7.** Intensity line profiles of emission and angular images from a silver nanowire for (**a**, **c**) the experimentally acquired data and (**b**, **d**) the simulated results. The top row shows the images,

and the bottom row shows the intensity line profiles extracted along the indicated green and orange arrows. Black arrows indicate the peak intensities.

The experimental results obtained within the detection bandwidth of 450 nm $\leq \lambda \leq$ 550 nm (Figure 5(e-g)) show good agreement with simulations performed at a single wavelength of $\lambda$ = 500 nm (Figure 5(h-j)), as discussed in the main text. Wavelength-integrated simulations were also carried out (Figure S8), which show a high degree of similarity to the single-wavelength case. This similarity originates from the small reflection phase pickup of the longitudinal SPP mode at the nanowire end facets. Using the CL wavelength-scan line profile shown in Figure 5(b), the reflection phase pickup $\delta$ can be extracted from the Fabry-Pérot cavity resonance condition $kL + \delta = m\pi$, assuming identical phase pickup at both facets. Here $k = \pi/d$ is the plasmon wavevector, $d$ is the spacing between neighboring standing-wave antinodes, $L$ is the nanowire length, and $m$ is the plasmon mode order. The average phase pickup for the $m$ = 4 mode across the wavelength bandwidth is $\delta_{\Delta\lambda}$ = 0.212π rad, which differs by only $\Delta\delta$ = 0.129π rad from the value obtained at λ = 500 nm, $\delta_{\lambda}$ = 0.083π rad. Consequently, the far-field interference over the 450-550 nm bandwidth is only weakly modified, leading to nearly identical emission patterns for the single-wavelength case (Figure 5(h-j)) and the wavelength-integrated case (Figure S8), consistent with the experimental results (Figure 5(e-g)).

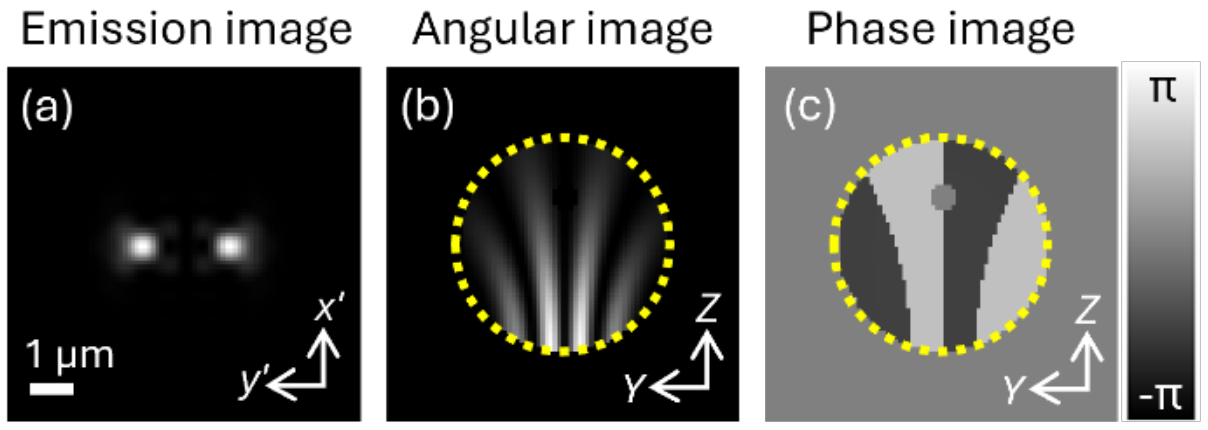

**Figure S8.** Wavelength-integrated (**a**) emission image, (**b**) angular image, and (**c**) phase image of the silver nanowire, calculated over the wavelength range 450 nm ≤ λ ≤ 550 nm. The yellow circle indicates the pinhole mask.

## Experimental methods

The CL measurements were performed using a modified STEM (JEM-2100F, JEOL, Japan) equipped with a spherical aberration corrector and a Schottky-type field emission gun. The instrument was operated at 80 kV with a 1 nA probe current and an illumination semi-angle of ~20 mrad, yielding a ~1 nm probe size[7]. CL emission was collected and collimated by a parabolic mirror before passing through a polarizer to select either *p*-polarized (*xz*-plane) or *s*-polarized (*y*-axis) electric field components. A beam splitter subsequently divided the polarized light into two parallel acquisition systems (see Figure 1(a) of the main text).

In the angle-resolved spectral system, the light passed through an angle-resolving pinhole mask (detection solid angle: 2.4 sr) positioned at emission angles $\theta = 20°$ and $\phi = 0°$ and was focused onto the spectrometer plane (Figure 1(a)). The system acquires three-dimensional (3D) datasets by raster scanning the electron beam across the sample in the *xy*-plane, generating spatial CL photon maps with each pixel contains a CL spectrum as a function of photon wavelength $\lambda$ at the fixed emission angles ($\theta$, $\phi$)[8, 9] (Figure 1(b)).

In the emission-position imaging system, two types of intensity datasets were recorded as pixelated camera images by toggling a diffraction lens (DL) (Figure 1(a)). Without the DL, the emitted light was recorded as emission spot(s) in the *x'y'*-plane of real space (i.e., emission space)[10, 11], as shown in the red-marked image in Figure 1(c). With the DL inserted, the system recorded the angular emission pattern, corresponding to angular space projected onto the *YZ*-plane, as shown in the blue-marked image in Figure 1(c), with the projected angular coordinates ($\theta$, $\phi$) also indicated[7]. To ensure consistency between both CL systems, identical 2.4 sr angle-resolving pinhole mask at ($\theta$, $\phi$) = (20°, 0°) was employed, as illustrated by the yellow circles in Figure 1(c). Bandpass filters were also used for mode selection.